\documentclass [12pt,a4paper]{article}

\usepackage{amssymb, theorem}
\newtheorem{lemma}{Lemma}

\newtheorem{thm}{Theorem}

\theorembodyfont{\rm}
\newtheorem{ex}{Example}
\def\Tr{{\rm Tr}\,}
\def\supp{{\rm supp}\,}
\def\ker{{\rm ker}\,}

\def\<{\langle}
\def\>{\rangle}
\def\qed{{\hfill $\square$}}
\begin{document}

\title{Quantum hypothesis testing and sufficient subalgebras}
\author{Anna Jen\v cov\'a\footnote{Email: jenca@mat.savba.sk. 
Supported by the Slovak Research and Development Agency
under the contract No. APVV 0071-06, grant VEGA  2/0032/09,
Center of Excellence SAS - Quantum Technologies
and ERDF OP R\&D Project CE QUTE ITMS  26240120009}
 \\ \\Mathematical Institute, Slovak Academy of Sciences\\
\v Stef\'anikova 49, 814 73 Bratislava, Slovakia}
\date{}
\maketitle

\begin{quote} We introduce a new notion of a sufficient subalgebra for
quantum states:  a subalgebra is 2- sufficient for a pair of states
$\{\rho_0,\rho_1\}$ if it contains all Bayes optimal tests of $\rho_0$ against
$\rho_1$. In classical statistics, this corresponds to the usual definition  of
sufficiency. We show this correspondence in the quantum setting for
some special cases. Furthermore, we show that sufficiency is equivalent to
2 - sufficiency, if the latter is required for $\{\rho_0^{\otimes n},\rho_1^{\otimes n}\}$, for all $n$.

\end{quote}

\begin{quote}
\end{quote}

\begin{quote}

\end{quote}

\section{Introduction}

In order to motivate our results, let us consider the following problem of 
classical statistics. Suppose that $P_0$ and $P_1$ are two probability distributions and the task is to discriminate between them by  an $n$-dimensional 
observation vector $X$. The problem is, if there is  a function (statistic)
$T: X\to Y$, such that the vector $Y=T(X)$ (usually of lower dimension) 
contains all information needed for the discrimination. 

In the setting of hypothesis testing, the  null
hypothesis $H_0=P_0$ is tested against
 the alternative $H_1=P_1$. In the most general formulation, a test  is a
 measurable
 function  $\varphi:X\to [0,1]$, which can be interpreted as the
  probability of rejecting the hypothesis if $x\in X$ occurs. 
There are two
   kinds of errors appearing in hypothesis testing: it may happen that $H_0$
 is rejected, although it is true (error of the first kind), or that it is not
 rejected when $H_1$  is true (error of the second kind). For a given test
 $\varphi$, the error probabilities are
 \begin{eqnarray*}
\alpha(\varphi)&=&\int \varphi(x)P_0(dx)\quad \mbox{ first kind}\\
\beta(\varphi)&=&\int (1-\varphi(x))P_1(dx)\quad \mbox{ second kind}
   \end{eqnarray*}

 The two kinds of errors are in some sense complementary and it is usually
  not possible to minimize both error probabilities simultaneously. In the
   Bayesian approach, we choose a prior probability distribution $\{\lambda, 1-\lambda\}$, $\lambda\in [0,1]$ on the two
  hypotheses and then minimize the average (Bayes) error probability
\[
 \int \varphi(x) \lambda P_0(dx)+ \int (1-\varphi(x))(1-\lambda)P_1(dx)=
 \lambda\alpha(\varphi)+(1-\lambda)\beta(\varphi).
\]

Suppose now that  $T$ is a sufficient statistic for $\{P_0, P_1\}$. 
Roughly speaking,  this means that there
exists a common version of the conditional expectation
$E[\cdot|T]= E_{P_0}[\cdot|T]$, $P_0$- a.s. and $E[\cdot|T]= E_{P_1}[\cdot|T]$, $P_1$- a.s. If $\varphi$ is any test, then $E[\varphi|T]$ is another test
 having the same error probabilities. It follows that we can always have an 
 optimal test that is a function of $T$, so that only values of 
 $T(X)$ are needed for optimal discrimination between $P_0$ and $P_1$.

The following theorem states that this can happen if and only if $T$ is sufficient, so that the above property characterizes sufficient 
statistics. The theorem was proved by Pfanzagl, see also \cite{strasser}.

\begin{thm}\label{thm:pfan}\cite{pfanzagl} Let $T:X\to Y$ be  a statistic. The following are equivalent.
  \begin{enumerate}
 \item For any $\lambda\in (0,1)$ and any test $\varphi:X\to [0,1]$, there exists a test $\psi:Y\to [0,1]$, such that
 \[
\lambda\alpha(\psi\circ T)+(1-\lambda)\beta(\psi\circ T)\le \lambda\alpha(\varphi)+(1-\lambda)\beta(\varphi)
 \]
\item $T$ is a sufficient statistic for $\{P_0,P_1\}$.
  \end{enumerate}

  \end{thm}

The problem of hypothesis testing can be considered also in the quantum setting.
Here we deal with a pair of density operators $\rho_0,\rho_1\in B(\mathcal H)$,
 where $\mathcal H$ is  a finite dimensional Hilbert space and all tests are
  given by operators $0\le M\le 1$, $M\in B(\mathcal H)$.
  The problem of
 finding the optimal tests (the quantum Neyman-Pearson tests) and average
 error probabilities was solved by Helstrom and Holevo \cite{helstrom,holevo}.

Here a question arises, if it is possible to discriminate the states 
optimally by measuring on a given subsystem.  Then we can gain some
information only on the restricted densities, which, in general, can be 
distinguished with less precision. 

Let
$M_0\subseteq B(\mathcal H)$ be the  subalgebra describing the subsystem we 
have access to. The average error probabilities for tests in 
$M_0$ are usually  higher than the optimal ones. 
We will consider the situation that this does not happen and 
  $M_0$ contains  some optimal tests for all prior probabilities.
In agreement with classical terminology (see \cite{strasser}), such a subalgebra will 
be called sufficient with respect to testing problems,
or 2-sufficient, for $\{\rho_0,\rho_1\}$.

The quantum counterpart of sufficiency was introduced  and studied
by Petz, see Chap. 9. in \cite{ohya&petz},
 in a more  general context.
According to this definition, the subalgebra $M_0$ is sufficient for
$\{\rho_0,\rho_1\}$, if there exists a completely positive, trace preserving map
 $M_0\to B(\mathcal H)$, that maps both restricted densities to the original
  ones. Then the restriction to $M_0$ preserves all information 
  needed for discrimination between the states and it is quite easy to see that
  a sufficient subalgebra must be
2-sufficient. 

The conditions for sufficiency seem to be quite restrictive (see for example 
the factorization conditions in \cite{ja&petz1}) and might be too strong, if only 
hypothesis testing is considered.
It is therefore natural to ask if there is a quantum version of Theorem
\ref{thm:pfan}, that is, if every 2-sufficient subalgebra must be sufficient.

In this paper, we give a partial answer to this question. We show that 
2-sufficiency and sufficiency are equivalent under each of the following conditions: 1) the subalgebra $M_0$ is
invariant under the modular group of one of the states, 2) 
$M_0$ is commutative, 3) $\rho_0$ and $\rho_1$ commute.
 Moreover, we show that if
the 2-sufficiency condition is strengthened to hold
for  $n$ independent copies of the densities for all $n$,
 then the two notions become equivalent.

The organization of the paper is as follows. In Section 2, some 
basic notions are introduced and several characterizations of a sufficient 
subalgebra are given. A new characterization, based on a version of the 
Radon-Nikodym derivative, is found, this will be needed for the main results.
Section 3 gives the quantum Neyman-Pearson lemma and quantum Chernoff bound.
Section 4 contains the main results: a convenient necessary condition for 
2-sufficiency is
 found and it is shown that it implies sufficiency in the three above described cases. Finally, the quantum Chernoff bound is utilized to treat the case when
  2-sufficiency holds for $n$ independent copies of the states, for all $n$.

\section{Some basic definitions and facts}

\subsection{Generalized conditional expectation}

Let $\mathcal H$ be a finite dimensional Hilbert space and let $\rho$
 be an invertible density matrix. Let $M_0\subseteq B(\mathcal H)$ be a subalgebra
 and let $E:B(\mathcal H)\to M_0$ be the trace preserving conditional
 expectation. Then $E(\rho)$ is the restricted density of the state
 $\rho$.

As we have seen, the classical sufficient statistic is defined by certain
property of the conditional expectations. It is well known that in the quantum
 case, a state preserving conditional expectation does not always exist.
Therefore we need the generalized conditional expectation, defined by 
Accardi and Cecchini \cite{accardi&cecchini}. In our setting, it can be given as follows.

Let us introduce the inner product $\<X,Y\>_\rho=\Tr X^* \rho^{1/2}Y\rho^{1/2}$
in $B(\mathcal H)$. Then the generalized conditional expectation $E_\rho$ is a
map $B(\mathcal H)\to M_0$, defined by
$$
\<X_0,Y\>_{\rho}=\<X_0,E_\rho(Y)\>_{E(\rho)},\quad X_0\in M_0, \ Y\in B(\mathcal H)
$$
It is easy to see that we have
\begin{equation}\label{eq:condexp}
E_\rho(X)=E(\rho)^{-1/2}E(\rho^{1/2}X\rho^{1/2})E(\rho)^{-1/2}
\end{equation}
It is known that $E_\rho$ is completely positive and unital and that
it is a conditional
expectation   if and only if $\rho^{it}M_0\rho^{-it}\subseteq M_0$, for all $t\in \mathbb R$. It is also
 easy to see that $E_\rho$ preserves the state $\rho$, that is,
 $E_\rho^*\circ E(\rho)=\rho$.

Next we introduce two subalgebras, related to  $E_\rho$.
Let $F_\rho$ be the set of fixed points of $E_\rho$ and
let $N_\rho\subseteq B(\mathcal H)$ be the multiplicative domain of $E_\rho$,
$$
N_\rho=\{X \in B(\mathcal H), E_\rho(X^*X)=E_\rho(X)^*E_\rho(X),
E_\rho(XX^*)=E_\rho(X)E_\rho(X)^*\}
$$
Then both $F_\rho$ and $N_\rho$ are  subalgebras  in $B(\mathcal H)$. It is clear that $F_\rho\subseteq M_0\cap N_\rho$, moreover, $X\in F_\rho$ if and only if
$\rho^{it}X\rho^{-it}\in M_0$ for all $t\in \mathbb R$. As  for
$N_\rho$, we have the following result.

\begin{lemma}\label{lemma:multiplicative}
$N_\rho=\rho^{1/2}M_0\rho^{-1/2}\cap \rho^{-1/2}M_0\rho^{1/2}$

\end{lemma}

{\it Proof.} It is clear from (\ref{eq:condexp}) that $X\in N_\rho$ if and only if
\begin{eqnarray*}
E(\rho^{1/2}X^*X\rho^{1/2})&=&E(\rho^{1/2}X^*\rho^{1/2})
E(\rho)^{-1}E(\rho^{1/2}X\rho^{1/2})\\
E(\rho^{1/2}XX^*\rho^{1/2})&=&E(\rho^{1/2}X\rho^{1/2})
E(\rho)^{-1}E(\rho^{1/2}X^*\rho^{1/2})
\end{eqnarray*}
Let $A=X\rho^{1/2}$, $B=\rho^{1/2}$. Similarly as in \cite{lieb&ruskai}, we put
$M=A-B\Lambda $, with $\Lambda=E(\rho)^{-1}E(\rho^{1/2}X\rho^{1/2})$. Then from
$E(M^*M)\ge 0$, we obtain
$$
E(A^*A)\ge E(A^*B)E(\rho)^{-1}E(B^*A),
$$
with equality if and only if $M=0$, this implies
$$\rho^{-1/2}X\rho^{1/2}=E(\rho)^{-1}E(\rho^{1/2}X\rho^{1/2})\in M_0.$$

Conversely, let $X_0=\rho^{-1/2}X\rho^{1/2}\in M_0$, then $E(\rho^{1/2}X\rho^{1/2})=E(\rho)X_0$, this implies that $M=0$.

Similarly, we get that $\rho^{-1/2}X^*\rho^{1/2}\in M_0$ is equivalent with the
second equality.

\qed

It is also known that $E_\rho(XY)=E_\rho(X)E_\rho(Y)$,
  $E_\rho(YX)=E_\rho(Y)E_\rho(X)$ for all $X\in N_\rho$, $Y\in B(\mathcal H)$,
   this can be also shown from the above Lemma.
Note that in the case that $E_\rho$ is a conditional expectation,
$F_\rho=N_\rho=M_0$.

\subsection{A Radon-Nikodym derivative and relative entropies}

Let $\rho_0,\rho_1$ be invertible density matrices in $B(\mathcal H)$. 
We will use the quantum version of the Radon-Nikodym derivative
 introduced in \cite{belavkin&staszewski}. In our setting, the derivative
$d_{\rho_0,\rho_1}$ of $\rho_1$ with respect to
$\rho_0$  is defined  as  the
 unique element  in $B(\mathcal H)$, such that
 $\Tr \rho_1X=\<X^*,d_{\rho_0,\rho_1}\>_{\rho_0}$. Then clearly
$$
d_{\rho_0,\rho_1}=\rho_0^{-1/2}\rho_1\rho_0^{-1/2}
$$
so that $d_{\rho_0,\rho_1}$ is positive, and $\|d_{\rho_0,\rho_1}\|\le \lambda$
 for any $\lambda>0$, such that $\rho_1\le \lambda\rho_0$. It is also easy to see that
$$
E_{\rho_0}(d_{\rho_0,\rho_1})=d_{E(\rho_0),E(\rho_1)}
$$

Let us recall that the Belavkin - Staszewski relative entropy
is defined as \cite{belavkin&staszewski}
$$
S_{BS}(\rho_1,\rho_0)=-\Tr\rho_0\eta(\rho_0^{-1/2}\rho_1\rho_0^{-1/2})=
-\Tr\rho_0\eta(d_{\rho_0,\rho_1})
$$
where $\eta(x)=-x\log(x)$. Let $S$ be the Umegaki relative entropy
$$
S(\rho_1,\rho_0)=\Tr \rho_1(\log \rho_1-\log\rho_0)
$$
then $S(\rho_1,\rho_0)\le S_{BS}(\rho_1,\rho_0)$, \cite{petz&hiai} and
$S(\rho_1,\rho_0)=S_{BS}(\rho_1,\rho_0)$ if
$\rho_0$ and $\rho_1$ commute.
Both relative entropies are monotone in the sense that
$$
S(\rho_1,\rho_0)\ge S(E(\rho_1),E(\rho_0)),\quad S_{BS}(\rho_1,\rho_0)\ge
S_{BS}(E(\rho_1),E(\rho_0))
$$
holds for any subalgebra $M_0$.
As we will see in the next section,  equality in the monotonicity for $S$ is
equivalent with sufficiency of  the subalgebra $M_0$ with respect to
$\{\rho_0,\rho_1\}$. For  $S_{SB}$, we have the following result.

\begin{lemma}\label{lemma:bs} The following are equivalent.
\begin{enumerate}
\item [(i)] $S_{BS}(\rho_1,\rho_0)=S_{BS}(E(\rho_1),E(\rho_0))$
\item [(ii)]  $d_{\rho_0,\rho_1}\in N_{\rho_0}$
\item[(iii)] $\rho_1\rho_0^{-1}\in M_0$
\item[(iv)]  $\rho_1\rho_0^{-1}=E(\rho_1)E(\rho_0)^{-1}$
\end{enumerate}
\end{lemma}

{\it Proof.} Since the function $-\eta(x)=x\log(x)$ is operator convex,
\begin{equation}\label{eq:jens}
\eta(d_{E(\rho_0),E(\rho_1)})=\eta(E_{\rho_0}(d_{\rho_0,\rho_1}))\le E_{\rho_0}(\eta(d_{\rho_0,\rho_1}))
\end{equation}
by Jensen's inequality. We have
$$
\Tr \rho_0  (E_{\rho_0}(\eta(d_{\rho_0,\rho_1}))-\eta(E_{\rho_0}(d_{\rho_0,\rho_1})))
=S_{BS}(\rho_1,\rho_0)-S_{BS}(E(\rho_1),E(\rho_0))
$$
and since $\rho_0$ is invertible, equality in the monotonicity of $S_{BS}$ is equivalent with equality in (\ref{eq:jens}). As it was proved in \cite{petzeq}, this
happens if and only if  $d_{\rho_0,\rho_1}\in N_{\rho_0}$.
This shows the equivalence  (i) $\leftrightarrow$ (ii). The equivalence of (ii)
and  (iii) follows by Lemma \ref{lemma:multiplicative}, (iii) $\iff$  (iv) is rather obvious.

\qed

\subsection{Sufficient subalgebras}

We say that the subalgebra $M_0\subseteq B(\mathcal H)$ is
sufficient for $\{\rho_0,\rho_1\}$ if there is a completely positive trace
preserving   map
$T: M_0 \to B(\mathcal H)$, such that $T\circ E(\rho_0)=\rho_0$ and
 $T\circ E(\rho_1)=\rho_1$. The following characterizations of
 sufficiency were obtained by Petz.

\begin{thm}\label{thm:suff} \cite{ja&petz,ohya&petz} The following are equivalent.
\begin{enumerate}
\item[(i)] $M_0\subseteq B(\mathcal H)$ is sufficient for $\{\rho_0,\rho_1\}$
\item[(ii)] $S(\rho_1,\rho_0)=S(E(\rho_1),E(\rho_0))$
\item[(iii)]  $\Tr \rho_0^s\rho_1^{1-s}=\Tr E(\rho_0)^sE(\rho_1)^{1-s}$ for
 some $s\in (0,1)$
 \item[(iv)] $\Tr E_{\rho_0}(X)\rho_1=\Tr X\rho_1 $ for all $X\in B(\mathcal H)$
\item[(v)] $E_{\rho_0}=E_{\rho_1}$.
\end{enumerate}
\end{thm}

The next characterization is based on the Radon-Nikodym
derivative.

\begin{thm}\label{thm:suffrn} The subalgebra $M_0\subseteq B(\mathcal H)$ is sufficient for $\{\rho_0,\rho_1\}$ if and only if   $d_{\rho_0,\rho_1}\in F_{\rho_0}$.

\end{thm}

{\it Proof.} Let us denote $d=d_{\rho_0,\rho_1}$ and
$d_0=d_{E(\rho_0),E(\rho_1)}$.
Since $d_0\in M_0$, we have by definition that
$$
\Tr \rho_1E_{\rho_0}(X)=\<d_0,E_{\rho_0}(X)\>_{E(\rho_0)}=\<d_0,X\>_{\rho_0}
$$
so  that $\Tr \rho_1E_{\rho_0}(X)=\Tr \rho_1X$ if and
 only if  $\<d_0,X\>_{\rho_0}=\<d,X\>_{\rho_0}$. It follows that $d=d_0$ is
 equivalent with sufficiency of $M_0$,  by Theorem \ref{thm:suff} (iv).
Since $E_{\rho_0}(d)=d_0$, this is equivalent with
 $d_{\rho_0,\rho_1}\in F_{\rho_0}$.

\qed

\section{Quantum hypothesis testing}

Let us now turn to the problem of hypothesis testing. Any test of the
hypothesis $H_0=\rho_0$ against the alternative $H_1=\rho_1$ is represented by
an operator $0\le M\le 1$, which corresponds to rejecting the hypothesis.
Then we have the error probabilities
\begin{eqnarray*}
&\alpha(M)=\Tr \rho_0M &\quad \mbox{ first kind}\\
&\beta(M)=\Tr \rho_1(1-M) &\quad \mbox{ second kind}
\end{eqnarray*}
For  $\lambda\in (0,1)$, we define  the Bayes optimal test to be a minimizer
of the expression
\begin{equation}\label{eq:bayes}
\lambda\alpha(M)+(1-\lambda)\beta(M)
\end{equation}
It is clear that minimizing (\ref{eq:bayes}) is the same as maximizing
$$
\Tr (\rho_1-t\rho_0)M,\qquad t=\frac{\lambda}{1-\lambda}
$$

\subsection{The quantum Neyman-Pearson lemma}

The following is the quantum version of the Neyman-Pearson lemma. The obtained
 optimal tests  are called the (quantum) Neyman-Pearson tests.
We give a simple proof  for completeness.

\begin{lemma}\label{lemma:qnp} Let $t\ge 0$ and let us denote
$P_{t,+}:=\supp (\rho_1-t\rho_0)_+$,
$ P_{t,-}:=\supp (\rho_1-t\rho_0)_-$ and $P_{t,0}:=1-P_{t,+}-P_{t,-}$.
Then the operator $0\le M_t\le 1$ is a Bayes optimal
test of $\rho_0$ against $\rho_1$ if and only if
$$
M_t=P_{t,+} +X_t
$$
where  $0\le X_t\le P_{t,0}$.
\end{lemma}

{\it Proof.} Let $0\le M\le 1$, then
\begin{eqnarray}
\Tr (\rho_1-t\rho_0)M&=&\Tr (\rho_1-t\rho_0)_+M-\Tr (\rho_1-t\rho_0)_-M\le\Tr (\rho_1-t\rho_0)_+M\nonumber\\
\label{eq:ef} &\le &
 \Tr (\rho_1-t\rho_0)_+=\Tr (\rho_1-t\rho_0)P_{t,+}
\end{eqnarray}
It follows that $M_t=P_{t,+}+X_t$, $X_t\le P_{t,0}$ is a Bayes optimal test.
Conversely, let $M_t$ be some Bayes optimal test, then we must have
$$
\Tr (\rho_1-t\rho_0)M_t=\Tr (\rho_1-t\rho_0)_+M_t=\Tr (\rho_1-t\rho_0)P_{t,+}
$$
so that $\Tr (\rho_1-t\rho_0)_-M_t=0$. By positivity, this implies that
$P_{t,-}M_t=M_tP_{t,-}=0$, so that
$$
M_t(P_{t,+}+P_{t,0})=(P_{t,+}+P_{t,0})M_t=M_t
$$
which is equivalent with $M_t\le P_{t,+}+P_{t,0}$. Furthermore, from
$$
\Tr(\rho_1-t\rho_0)_+(P_{t,+}+P_{t,0}-M_t)=0
$$
we obtain $P_{t,+}-P_{t,+}M_tP_{t,+}=P_{t,+}(1-M_t)P_{t,+}=0$, hence
$(1-M_t)P_{t,+}=0$. We obtain $P_{t,+}\le M_t$ and by
putting $X_t:=M_t-P_{t,+}$, we get the result.

\qed

Let us denote by $\Pi_{e,\lambda}$ the minimum Bayes error probability. Then
\begin{eqnarray}
\Pi_{e,\lambda}&=&\lambda \alpha(M_{\lambda/(1-\lambda)})+(1-\lambda)\beta(M_{\lambda/(1-\lambda)})=\nonumber\\
&=&\frac12(1-\|(1-\lambda)\rho_1-\lambda\rho_0\|_1)\label{eq:minprob}
\end{eqnarray}
where the last equality follows from
\[
1-t =\Tr (\rho_1-t\rho_0)=\Tr (\rho_1-t\rho_0)_+-\Tr(\rho_2-t\rho_0)_-
\]
and
\[
\|\rho_1-t\rho_0\|_1=\Tr |\rho_1-t\rho_0|=\Tr (\rho_1-t\rho_0)_++\Tr(\rho_2-t\rho_0)_-
\]

\subsection{The quantum Chernoff bound}

Suppose now that we have $n$ copies of the states $\rho_0$ and $\rho_1$,
 so that we test the hypothesis $\rho_0^{\otimes n}$ against
 $\rho_1^{\otimes n}$ by means of an operator $0\le M_n\le 1$,
 $M_n\in \mathcal B(\mathcal H^{\otimes n})$. Again, we may use the
 Neyman-Pearson lemma to find the minimum Bayes error probability
 $$
\Pi_{e,\lambda,n}=\frac12(1-\|(1-\lambda)\rho_1^{\otimes n}-\lambda\rho_0^{\otimes n}\|_1)
 $$
The following important result, obtained in \cite{chernoff} and \cite{nussbaum&szkola} (see also \cite{chernoffall}), is the quantum
version of the classical Chernoff bound:
\begin{equation}\label{eq:chernoff}
\lim_n (-\frac 1n\log \Pi_{e,\lambda,n})=-\log(\inf_{0\le s\le 1}\Tr \rho_0^{1-s}\rho_1^s)=: \xi_{QCB}(\rho_0,\rho_1)
\end{equation}
The expression $\xi_{QCB}$ has a number of interesting
properties. For example, it was proved that it is always nonnegative and
equal to 0 if and only if
 $\rho_0=\rho_1$, moreover, it is monotone in the sense that
 \[
\xi_{QCB}(\rho_0,\rho_1)\ge \xi_{QCB}(E(\rho_0),E(\rho_1))
 \]
Therefore, although it is not symmetric, $\xi_{QCB}$ provides a reasonable
distance measure on density matrices, called the quantum Chernoff distance.
Note also that in the case that the matrices are invertible, the infimum
  is always attained in some $s^*\in[0,1]$.

\section{2-sufficiency}

 We say that $M_0$ is
sufficient with respect to testing problems, or 2-sufficient,
for $\{\rho_0,\rho_1\}$ if for any test $M$ and any $\lambda\in (0,1)$,
there is  some test $N_\lambda\in M_0$, such that
$$
\lambda\alpha(N_\lambda)+(1-\lambda)\beta(N_\lambda)\le \lambda\alpha(M)+
(1-\lambda)\beta(M)
$$
It is quite clear that $M_0$ is 2-sufficient if and only if for all $t\ge 0$,
 we can find a Neyman-Pearson test $M_t\in M_0$. Moreover, suppose that $M_0$
  is a sufficient subalgebra for $\{\rho_0,\rho_1\}$ and let $T=E_{\rho_0}=
  E_{\rho_1}$. Then, if $M_t$ is a Neyman-Pearson test, then $T(M_t)\in M_0$
  is a Neyman-Pearson test as well. Hence, a sufficient subalgebra is
  always 2-sufficient. In this section, we find the opposite implication in 
  some special  cases.

\begin{lemma}\label{lemma:eigen} $P_{t,0}\ne 0$ if and only if $t$ is an eigenvalue of $d:=d_{\rho_0,\rho_1}$.
Moreover, the rank of $P_{t,0}$ is equal to multiplicity of $t$.
\end{lemma}

{\it Proof.} By definition,
$$
(\rho_1-t\rho_0)P_{t,0}=\rho_0^{1/2}(d-t)\rho_0^{1/2}P_{t,0}=0
$$
so that $(d-t)\rho_0^{1/2}P_{t,0}\rho_0^{1/2}=0$. Suppose $P_{t,0}\ne 0$,
then  $t$ is an eigenvalue of $d$ and any vector in the  range of
$\rho_0^{1/2}P_{t,0}\rho_0^{1/2}$ is an eigenvector.
This implies that $r(P_{t,0})=r(\rho^{1/2}P_{t,0}\rho^{1/2})\le r(F)$, where
 $F$ is the eigenprojection of $t$.

Conversely, let $t$ be an eigenvalue of $d$
with the  eigenprojection $F$, then
$$
(\rho_1-t\rho_0)\rho_0^{-1/2}F\rho_0^{-1/2}=\rho_0^{1/2}(d-t)F\rho_0^{-1/2}=0,
$$
so that the range of   $\rho^{-1/2}F\rho^{-1/2}$ is in the kernel of
$\rho_1-t\rho_0$, this implies $r(F)\le r(P_{t,0})$.

\qed

Let us denote $Q_{t,+}=\supp (E(\rho_1)-tE(\rho_0))_+$, $Q_{t,0}=\ker (E(\rho_1)-tE(\rho_0))$ and let $\Pi_{e,\lambda}^0$ be the minimal Bayes error probability
for the restricted densities
\[
\Pi_{e,\lambda}^0:=\inf_{M\in M_0}\lambda\alpha(M)+(1-\lambda)\beta(M)=
\frac12(1-\|(1-\lambda)E(\rho_1)-\lambda E(\rho_0)\|_1)
\]

\begin{lemma}\label{lemma:projs} The following are equivalent.
\begin{enumerate}
\item[(i)] The subalgebra $M_0$ is 2-sufficient for
$\{\rho_0,\rho_1\}$.
\item[(ii)] $\Pi_{e,\lambda}^0=\Pi_{e,\lambda}$ for all $\lambda\in (0,1)$.
\item[(iii)] $Q_{t,0}=P_{t,0}$ and $Q_{t,+}=P_{t,+}$ for all $t\ge 0$.

\end{enumerate}
\end{lemma}

{\it Proof.}   It is obvious that (i) implies (ii). Suppose (ii) and let
 us denote $f(t):=\max_{0\le M\le 1} \Tr (\rho_1-t\rho_0)M$. If $N_t$ is any
 Neyman-Pearson test for $\{E(\rho_0), E(\rho_1)\}$, then
\[
\Tr (\rho_1-t\rho_0)N_t=\Tr (E(\rho_1)-tE(\rho_0))N_t= f(t),
\]
 so that
$N_t$ is a Neyman-Pearson test for $\{\rho_0,\rho_1\}$ as well.
Putting $N_t=Q_{t,+}$ and $N_t=Q_{t,+}+Q_{t,0}$, we get by Lemma
\ref{lemma:qnp} that
\[
Q_{t,+}=P_{t,+}+X_t,\qquad Q_{t,+}+Q_{t,0}=P_{t,+}+Y_t,
\]
  with $X_t,Y_t\le P_{t,0}$. This implies that $Q_{t,0}\le P_{t,0}$ and
  $Q_{t,+}=P_{t,+}$ if $P_{t,0}=0$.

Let $t$ be an eigenvalue of $d_0$, then
 $P_{t,0}\ge Q_{t,0}\ne  0$, hence $t$ is also an eigenvalue of $d$, and its multiplicity
 in $d_0$ is not greater  that its multiplicity in $d$. Since the sum of
 multiplicities
 must equal to $m=\dim(\mathcal H)$, we must have $r(Q_{t,0})=r(P_{t,0})$, so
 that $Q_{t,0}=P_{t,0}$. This implies that $X_t\le Q_{t,0}$, hence $X_t=0$
 and $P_{t,+}=Q_{t,+}$ for all $t$.

The implication (iii) $\rightarrow$ (i)  is again obvious.

 \qed

Note that the condition (ii) is equivalent with 
\[
\|E(\rho_1)-t E(\rho_0)\|_1\ge\|\rho_1-t\rho_0\|_1,\quad \mbox{for all } t\ge 0
\]
This condition, with $E(\rho_0)$ and $E(\rho_1)$ replaced by arbitrary densities $\sigma_0$ and $\sigma_1$ was studied in \cite{alberti&uhlmann}. 
It was shown
that for $2\times 2$ matrices, this is equivalent with the existence of a 
completely positive trace preserving map $T$, such that $T(\rho_0)=\sigma_0$ and
 $T(\rho_1)=\sigma_1$. In 
 our case, this means that 2-sufficiency implies sufficiency for $2\times 2$ 
 matrices. Since
  any nontrivial subalgebra in $\mathcal M(\mathbb C^2)$ is commutative, this agrees with our results below.

The above Lemma gives  characterizations of 2-sufficiency, but the conditions
are not easy  to check.  The next Theorem gives a simple necessary condition.

\begin{thm}\label{thm:2suff} Let $M_0$ be 2-sufficient for $\{\rho_1,\rho_0\}$. Then
$d_{\rho_1,\rho_0}\in N_{\rho_0}$.
\end{thm}

{\it Proof.} By the previous Lemma, we have $P_{t,0}=Q_{t,0}\in M_0$ for all $t$.
Let  $t_1,\dots,t_k$ be the eigenvalues of $d$ and denote $P_i=P_{t_i,0}$. Then
from $(d-t_i)\rho_0^{1/2}P_i=0$ we get
$$
d\rho_0^{1/2}\sum_iP_i=\rho_0^{1/2}\sum_it_iP_i
$$
By Lemma \ref{lemma:eigen} and its proof,
$\supp (\rho_0^{1/2}P_i\rho_0^{1/2})\le F_i$
 and $r(P_i)=r(F_i)$, with $F_i$ the eigenprojection of $t_i$.
It follows that $\sum_i\rho_0^{1/2}P_i\rho_0^{1/2}$, and hence also $\sum_iP_i$,
is invertible. Therefore,
$$
d\rho_0^{1/2}=\rho_0^{1/2}c,\qquad
c:=\sum_it_iP_i(\sum_j P_j)^{-1}
$$
that is, $d=\rho_0^{1/2}c\rho_0^{-1/2}$, with $c\in M_0$. Moreover,
$d=d^*=\rho_0^{-1/2}c^*\rho_0^{1/2}$, so that $d\in \rho_0^{1/2}M_0\rho_0^{-1/2}\cap
\rho_0^{-1/2}M_0\rho_0^{1/2}$. By Lemma \ref{lemma:multiplicative}, this entails
 that $d\in N_{\rho_0}$.

 \qed

\begin{thm} Let the subalgebra $M_0$ be  2-sufficient for $\{\rho_0,\rho_1\}$.
Then $M_0$ is sufficient for $\{\rho_0,\rho_1\}$ in each of the following cases.
\begin{enumerate}
\item [(1)] $\rho_0^{it}M_0\rho_0^{-it}
\subseteq M_0$ for all $t\in \mathbb R$ 
\item [(2)] $M_0$ is commutative
\item [(3)] $\rho_0$ and $\rho_1$ commute
\end{enumerate}
\end{thm}

{\it Proof.} (1) By Theorem \ref{thm:2suff}, we have $d\in N_{\rho_0}$. 
Since $\rho_0^{it}M_0\rho_0^{-it}\subseteq M_0$, we have 
$d\in N_{\rho_0}=F_{\rho_0}$. 
 By Theorem \ref{thm:suffrn}, this implies that $M_0$ is sufficient.

(2) Since $d\in N_{\rho_0}$, we have $S_{BS}(\rho_1,\rho_0)=
S_{BS}(E(\rho_1),E(\rho_0))$, by Lemma \ref{lemma:bs}. Since
$M_0$ is commutative,
$$
S(E(\rho_1),E(\rho_0))=S_{BS}(E(\rho_1),E(\rho_0))=S_{BS}(\rho_1,\rho_0)\ge S(\rho_1,\rho_0)
$$
By monotonicity of the relative entropy, this implies $S(\rho_1,\rho_0)=
S(E(\rho_1),E(\rho_0))$, so that $M_0$ is sufficient for $\{\rho_0,\rho_1\}$,
 by Theorem \ref{thm:suff} (ii).

(3) Let $M_1$ be the subalgebra generated by all $P_{t,+}$,
 $t\in \mathbb R$. Then $M_1$ is commutative and 2-sufficient for
 $\{\rho_0,\rho_1\}$, hence sufficient by (2).
 If $M_0$ is 2-sufficient, we must have
 $M_1\subseteq M_0$ by Lemma \ref{lemma:projs}, so that  $M_0$ must be
 sufficient for $\{\rho_0,\rho_1\}$ as well.

 \qed

It is clear from the proof of (1) that 2-sufficiency implies sufficiency whenever 
$N_{\rho_0}=F_{\rho_0}$ (or, equivalently, $N_{\rho_1}=F_{\rho_1}$). In fact, it 
can be shown that $N_{\rho_0}=F_{\rho_0}$ whenever 
$M_0$ is commutative, which gives an alternative proof of (2). Next we give a 
further example of this situation.

\begin{ex} Let $\mathcal H=\mathbb C^4$ and let 
$M_0=\mathcal M(\mathbb C^2)\otimes I\subset B(\mathcal H)$. Let $\rho$ be a 
block-diagonal density matrix $\rho=\left( \begin{array}{cc} \rho_1 & 0\\
  0 & \rho_2 \end{array}\right)$, where $\rho_1,\rho_2$ are positive invertible matrices in $\mathcal M(\mathbb C^2)$, and let $\sigma$ be any density matrix.
  Suppose that $M_0$ is 2-sufficient for $\{\rho,\sigma\}$.

By Theorem \ref{thm:2suff}, $d_{\sigma,\rho}\in N_\rho$, which by Lemma 
\ref{lemma:bs} is equivalent with $\sigma\rho^{-1}\in M_0$. This implies that
$\sigma$ must be  block-diagonal as well, $\sigma=\left( \begin{array}{cc} 
\sigma_1 & 0\\
  0 & \sigma_2 \end{array}\right)$.

By Lemma \ref{lemma:projs}, $P_{t,+}\in M_0$ for all $t\ge 0$, so that
$P_{t,+}=\left( \begin{array}{cc} p_t & 0\\
  0 & p_t \end{array}\right)$, where $p_t=\supp (\sigma_1-t\rho_1)_+=
  \supp(\sigma_2-t\rho_2)_+$. Since $p_t$ is a projection in $\mathcal M(\mathbb C^2)$, we have he following two possibilities: either $p_t=I$ for $t<t_0$ and
   $p_t=0$ for $t\ge t_0$, or $p_t$ is one-dimensional for $t$ in some interval $(t_0,t_1)$. 
   Since $\rho=\sigma$ in the first case, we may suppose that the latter is 
   true, so that $p_t$ is a common eigenprojection of $\sigma_1-t\rho_1$ and
   $\sigma_2-t\rho_2$ for $t\in (t_0,t_1)$. It follows that $\sigma_1-t\rho_1$
 commutes with $\sigma_2-t\rho_2$ for $t\in (t_0,t_1)$, which implies that $\rho_1$ commutes with $\rho_2$.  

Let $X\in N_\rho$, then 
$X= \rho^{1/2} X_0\rho^{-1/2}$, where both $X_0$, $\rho X_0 \rho^{-1}\in M_0$.
Let $X_0=Y\otimes I\in M_0$, then $\rho X_0\rho^{-1}\in M_0$ if and only if
$\rho_1 Y\rho_1^{-1}=\rho_2 Y \rho_2^{-1}$, that is, $Y$ commutes with 
$\rho_2^{-1}\rho_1$. If $\rho_2^{-1}\rho_1$ is a constant, then $\rho^{it}M_0
\rho^{-it}\subseteq M_0$, so that $F_\rho=M_0=N_\rho$. Otherwise, $Y$ must 
commute with both $\rho_1$ and $\rho_2$ and in this case, $X=\rho^{1/2}X_0\rho^{-1/2}=X_0\in F_\rho$.

In conclusion, if $M_0$ is 2-sufficient for $\{\rho,\sigma\}$, we must have
$N_\rho=F_\rho$, so that $M_0$ must be a sufficient subalgebra.
\qed

\end{ex}

Let us now suppose that we have $n$ independent copies of the states, 
$\rho_0^{\otimes n}$ and $\rho_1^{\otimes n}$. An optimal test for  $H_1: \rho_0^{\otimes n}$ against $H_1:\rho_1^{\otimes n}$  usually cannot be obtained as
 the product of optimal tests, but we may  ask if there is some optimal test in 
 $M_0^{\otimes n}$. If this is the case for all $\lambda$, we say that $M_0$ 
 is $(2,n)$-sufficient  for $\{\rho_0,\rho_1\}$.

\begin{thm} The following conditions are equivalent.
\begin{enumerate}
\item [(i)]  $M_0$ is
$(2,n)$-sufficient for $\{\rho_0,\rho_1\}$, for all $n$.
\item [(ii)] $M_0$ is a sufficient subalgebra for $\{\rho_0,\rho_1\}$.
\end{enumerate}
\end{thm}

{\it Proof.} Let us denote
$$
\Pi^0_{e,\lambda, n}:=\frac12(1-\|(1-\lambda)E(\rho_1)^{\otimes n}-\lambda E(\rho_0)^{\otimes n}\|_1)
$$
By Lemma \ref{lemma:projs} (ii),
the condition (i) implies that $\Pi_{e,\lambda,n}=\Pi^0_{e,\lambda, n}$
for all $n$, hence also
\[
\lim_n(-\frac 1n\log \Pi_{e,\lambda,n})=\lim_n(-\frac 1n\log \Pi^0_{e,\lambda,
n})
\]
By (\ref{eq:chernoff}), this entails that
$$
\inf_{0\le s\le 1} \Tr \rho_0^{1-s}\rho_1^s=\inf_{0\le s\le 1} \Tr E(\rho_0)^{1-s}E(\rho_1)^s
$$
By monotonicity, we have $\Tr \rho_0^{1-s}\rho_1^s\le \Tr E(\rho_0)^{1-s}E(\rho_1)^s$ for all $s\in [0,1]$. Suppose that the infimum on the RHS is attained in
 some $s_0\in [0,1]$. Then
$$
\Tr E(\rho_0)^{1-s_0}E(\rho_1)^{s_0}=\inf_{0\le s\le 1} \Tr \rho_0^{1-s}\rho_1^s\le \Tr\rho_0^{1-s_0}\rho_1^{s_0}.
$$
If $s_0=0$ or 1, then the quantum Chernoff distance is equal to 0, so that
$\rho_0=\rho_1$ and the subalgebra $M_0$ is trivially sufficient. Otherwise,
we must have
$\Tr E(\rho_0)^{1-s_0}E(\rho_1)^{s_0}=\Tr\rho_0^{1-s_0}\rho_1^{s_0}$ for
$s_0\in (0,1)$, which implies that $M_0$ is sufficient for $\{\rho_0,\rho_1\}$,
 by Theorem \ref{thm:suff} (iii).

Conversely, let $E_{\rho^{\otimes n}}$ be the generalized conditional expectation $B(\mathcal H^{\otimes n}) \to
M_0^{\otimes n}$. It is easy to see that for any invertible density matrix
$\rho$,
$E_{\rho^{\otimes n}}=E_{\rho}^{\otimes n}$, so that if $E_{\rho_0}=E_{\rho_1}$,
 then $E_{\rho_0^{\otimes n}}=E_{\rho_1^{\otimes n}}$ for all $n$.
Hence if $M_0$   is sufficient for
$\{\rho_0, \rho_1\}$, then $M_0^{\otimes n}$ is sufficient for
$\{\rho_0^{\otimes n},\rho_1^{\otimes n}\}$ for all $n$, this implies (i).

\qed

\end{document}